\documentclass[aps,prb,superscriptaddress,twocolumn,preprintnumbers,amsmath,amssymb]{revtex4-1}
\usepackage{color}
\usepackage{array}
\usepackage{amsmath}
\usepackage{amssymb}
\usepackage{graphicx}
\usepackage{epstopdf}
\usepackage{hyperref}
\usepackage{float}
\usepackage{bibentry}
\usepackage{tcolorbox}
\usepackage{bm}
\usepackage{dsfont}
\usepackage[scr=boondoxo]{mathalfa}
\usepackage{xcolor}
\usepackage{braket}

\newcommand{\ri}{\mathrm{i}}

\begin{document}
\title{Nonreciprocal directional dichroism induced by a temperature gradient as a probe for mobile spin dynamics in quantum magnets}
\author{Xu Yang}
\affiliation{Department of Physics, Boston College, Chestnut Hill, MA 02467, USA}
\author{Ying Ran}
\affiliation{Department of Physics, Boston College, Chestnut Hill, MA 02467, USA}
\date{\today}
\begin{abstract}
Novel states of matter in quantum magnets like quantum spin liquids attract considerable interest recently. Despite the existence of a plenty of candidate materials, there is no confirmed quantum spin liquid, largely due to the lack of proper experimental probes. For instance, spectrosocopy experiments like neutron scattering receive contributions from disorder-induced local modes, while thermal transport experiments receive contributions from phonons. Here we propose a thermo-optic experiment which directly probes the mobile magnetic excitations in spatial-inversion symmetric and/or time-reversal symmetric Mott insulators: the temperature-gradient-induced nonreciprocal directional dichroism (TNDD) spectroscopy. Unlike traditional probes, TNDD directly detects mobile magnetic excitations and decouples from phonons and local magnetic modes.
\end{abstract}
\maketitle
\textbf{Introduction}
Quantum spin liquids(QSL), proposed by Anderson\cite{ANDERSON1973153} for spatial dimensions $> 1$, attracted considerable interest in the past decades (see Ref.\cite{Lee_2014,RevModPhys.89.025003,Savary_2016,Broholmeaay0668} for reviews). Although theoretically these novel states of matter are known to exist and have even been successfully classified\cite{PhysRevB.65.165113,RevModPhys.89.041004}, to date there is no experimentally confirmed QSL material. As a matter of fact, an increasing list of candidate QSL materials emerges recently due to the extensive experimental efforts, including, for instance, Herbertsmithite\cite{doi:10.1021/ja053891p,RevModPhys.88.041002}, $\alpha$-RuCl$_3$ under a magnetic field\cite{Kasahara2018}, and quantum spin ice materials\cite{PhysRevB.69.064404,Gingras_2014}. An outstanding challenge in this field is the lack of appropriate experimental probes. Traditional probes for magnetic excitations include thermodynamic measurements, various spectroscopy measurements such as neutron scattering and nuclear magnetic resonance, and the thermal transport. Ideally, one would like to directly probe the mobile magnetic excitations in a QSL, such as the fractionalized spinons. The major limitation of traditional probes is from the contributions of other degrees of freedom; e.g., the spectroscopy measurements couple to local impurity modes, while the thermal transport couple to phonons. It is highly nontrivial to directly probe the intrinsic contribution from the mobile magnetic excitations. To highlight this challenge, there is no known direct probe to even detect the mobility gap of magnetic excitations, which is fundamentally important in the field of topologically ordered states.

In this paper we propose a thermo-optic experiment which serves as a new probe for mobile magnetic excitations in Mott insulators respecting either the spatial inversion symmetry $\mathcal{I}$ or the time-reversal symmetry $\mathcal{T}$ \footnote{or time-reversal symmetry combined with a spatial translation such as in an antiferromagnet}, or both: the temperature-gradient-induced nonreciprocal directional dichroism (TNDD). In a sense TNDD combines the thermal transport and optical spectroscopy together, and effectively decouples from phonon and local magnetic modes.

\textbf{Theory of TNDD} Nonreciprocal directional dichroism (NDD) is a phenomenon referring to the difference in the optical absorption coefficient between counterpropagating lights\cite{doi:10.1080/14786436508224252}. From the Fermi's golden rule, NDD for linearly polarized lights is due to the interference between the electric dipole and magnetic dipole processes\cite{PhysRevB.89.184419}\footnote{In general NDD receives contributions from higher order multipole processes. \cite{PhysRevLett.122.227402} However in the context of Mott insulators the electric-dipole-magnetic-dipole contribution Eq.\ref{eq:tndd_micro} dominates.}:
\begin{align}
 &\delta_{\hat n}\alpha(\omega)\equiv \alpha_{\hat n}(\omega)-\alpha_{-\hat n}(\omega)=\frac{2\mu_r}{\epsilon_0 c^2}\frac{2\pi}{\hbar}\cdot\frac{\hbar\omega}{V} \sum_{i,f} (\rho_i-\rho_f) \cdot 2\notag\\
& \cdot \mbox{Re[}\langle i|\vec P\cdot \hat{\mathcal{E}}|f \rangle\langle f|\vec M\cdot \hat{\mathcal{B}}|i\rangle]\cdot\delta(E_f-E_i-\hbar \omega)\label{eq:tndd_micro}
\end{align}
where $\alpha_{\pm\hat n}(\omega)$ is the optical absorption coefficient of counterpropagating lights (along $\pm \hat n$) at frequency $\omega$, which are $\mathcal{I}$ (or $\mathcal{T}$) images of each other. $\vec P$ ($\vec M$) is the electric polarization (magnetic moment) operator. $\hat{\mathcal{E}}$ ($\hat{\mathcal{B}}$) is the direction of the electric field (magnetic field) and $\hat n\sim \hat{\mathcal{E}}\times \hat{\mathcal{B}}$. $|i\rangle,|f\rangle$ label the initial and final states in the optical transition ($\rho_i$ and $\rho_f$ are their density matrix elements), $\epsilon_0$ and $c$ are the vacuum permittivity and the speed of light, $V$ is the volume of the material, and $\mu_r$ is the material's relative permeability. Clearly both $\mathcal{I}$ and $\mathcal{T}$ need to be broken to have a nonzero NDD because $\mbox{Re}[\langle i|P|f\rangle\cdot \langle f|M|i\rangle]$ is odd under either symmetry operation. NDD has been actively applied in the field of multiferroics\cite{PhysRevLett.85.4385,PhysRevLett.92.137401,Arima_2008,PhysRevLett.106.057403,Takahashi2012,Okamura2013,Kezsmarki2014,PhysRevLett.115.267207,Tokura2018} to probe the dynamical coupling between electricity and magnetism.

\begin{figure}
 \includegraphics[width=0.45\textwidth]{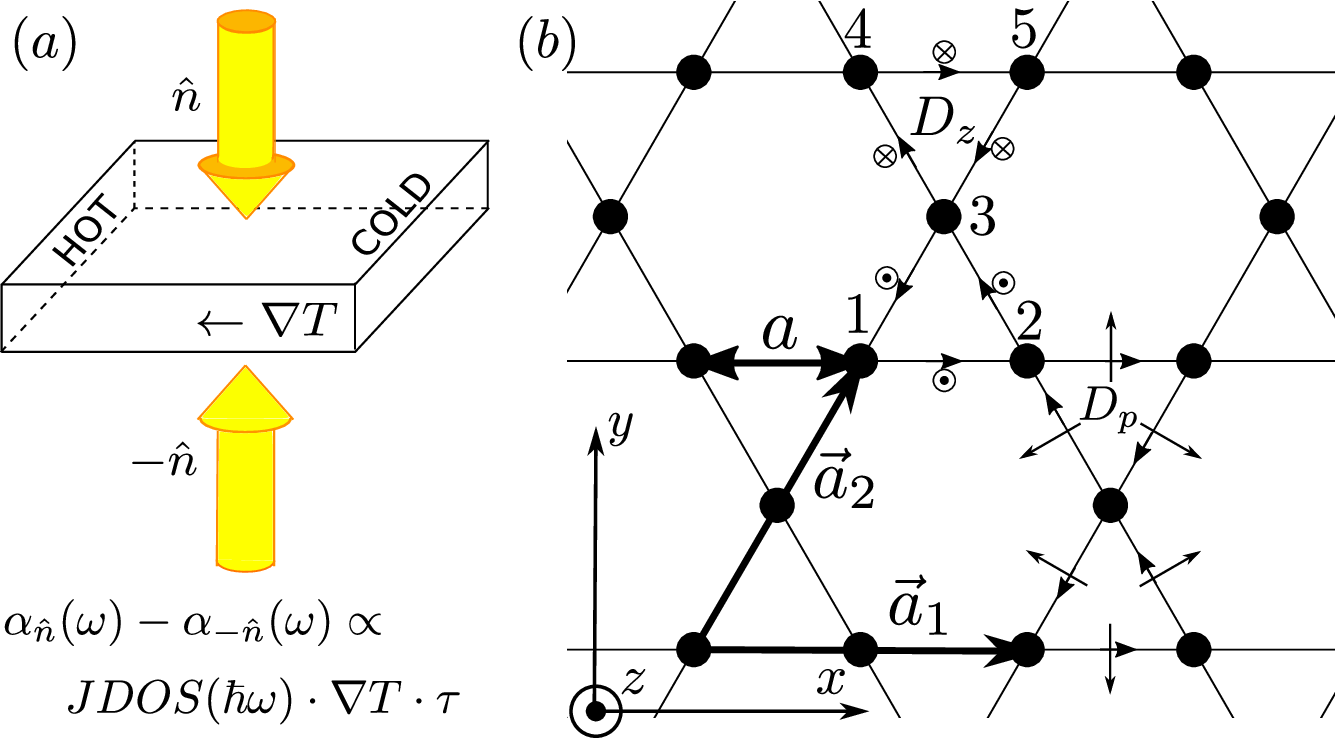}
 \caption{(a): A schematic illustration of the TNDD effect: in the presence of a temperature gradient, the optical absorption coefficients for counterpropagating lights become different, which essentially probes the joint density of states of mobile magnetic excitations. (b) A Kagome lattice and the Dzyaloshinskii–Moriya vectors $\vec D_{ij}$.}
 \label{fig:DM_terms}
\end{figure}

The TNDD spectroscopy essentially detects the joint density of states of mobile magnetic excitations, and can be intuitively understood as follows (see Fig.\ref{fig:DM_terms}(a)). Consider a Mott insulator respecting $ \mathcal{I}$ and/or $\mathcal{T}$ so that NDD vanishes in thermal equilibrium. In the presence of a temperature gradient, the system reaches a nonequilibrium steady state with a nonzero heat current carried by mobile excitations. For simplicity one may assume that excitations of the system are well-described by quasiparticles, e.g., spinons or magnons, phonons, etc. The leading order nonequilibrium change of $\rho_i$ and $\rho_f$ in Eq.(\ref{eq:tndd_micro}) satisfies $\delta \rho_i,\delta\rho_f\propto \nabla T\cdot \tau$ from a simple Boltzmann equation analysis, where $\tau$ is the relaxation time. 

\emph{The crucial observation is that this nonequilibrium state breaks both the inversion symmetry (by $\nabla T$) and the time-reversal symmetry (by $\tau$).} Consequently one expects a NDD signal proportional to $\nabla T\cdot \tau$. Precisely speaking TNDD is a second-order thermo-electromagnetic nonlinear response: it is a change of optical absorption (a linear response) due to a temperature gradient. The factor $\nabla T\cdot \tau$ in TNDD indicates that it is a generalization of Drude-phenomenon to nonlinear responses. Notice that the Drude-phenomenon is independent of whether the system has a quasiparticle description or not. Even in the absence of quasiparticle descriptions, strongly interacting liquids may have nearly conserved momentum. The relaxation time $\tau$ in Drude physics should be interpreted as the momentum relaxation time\cite{PhysRevB.75.245104}. This indicates that TNDD discussed here can be generalized to systems without quasiparticle descriptions such as the U(1)-Dirac spin liquid\cite{PhysRevB.37.3774,PhysRevB.70.214437,PhysRevLett.98.117205} and the spinon Fermi surface state\cite{PhysRevB.72.045105,PhysRevLett.95.036403}.

\textbf{Advantages of TNDD spectroscopy}
Now we comment on the major advantages of TNDD as a probe of spin dynamics. First, TNDD is a dynamical spectroscopy with the frequency resolution in contrast to the DC thermal transport, and essentially probes the joint density of states of magnetic excitations. Second, the fact that TNDD only receives contributions from $\mbox{Re}[\langle i|P|f\rangle\cdot \langle f|M|i\rangle]$ dictates that the \emph{phonons}' contribution can be safely ignored: The natural unit for the magnetic moment of phonon, the nuclear magneton, is more than three orders of magnitudes smaller than that of the electron, the Bohr magneton. 

In addition, at the intuitive level, a local magnetic mode (e.g. from a magnetic impurity atom) can only couple to a local temperature instead of a temperature gradient. A local temperature respects both $\mathcal{I}$ (after taking disorder-average) and $\mathcal{T}$. Consequently, such local modes are not expected to contribute to TNDD either. From a more careful estimate (see App.\ref{app:localized_modes} for detailed discussions), we find that the contribution to TNDD from localized modes with a localization length $\xi$, comparing to the contribution from the intrinsic mobile magnetic modes, is at least down by a factor of $\xi/l_m$, where $l_m$ is the mean-free path of the mobile magnetic excitations. We have assumed that $\xi\ll l_m$: for local magnetic modes carried by magnetic impurity atoms or crystalline defects, typically $\xi$ is comparable with the lattice spacing $a$, while usually $l_m\gg a$ in a reasonably clean Mott insulator at low temperatures. 

\textbf{Estimate of the TNDD response}
One may estimate the size of TNDD signal in a spin-orbital coupled Mott insulator. The relevant dimensionless ratio limiting the experimental resolution is: 
\begin{align}
TNDD(\omega)\equiv \frac{\delta_{\hat n}\alpha(\omega)}{\alpha_{\hat n}(\omega)+\alpha_{-\hat n}(\omega)}.\label{eq:TNDD_ratio}
\end{align} 
In a Mott insulator, the polarization carried by a magnetic excitation can be estimated as $\zeta \cdot e\cdot a$, where $a$ is the lattice spacing and $\zeta$ is dimensionless. Assuming the average temperature of the system $k_B T$ to be comparable to the magnetic excitation energy\footnote{Similar to a thermal transport experiment, if the temperature of the system is far below the magnetic excitation energy, a temperature gradient would not efficiently affect the magnetic excitation distributions and would not lead to a sizable TNDD.}, we find that (see App.\ref{app:dm} for details):
\begin{align}
 TNDD(\omega)\sim \left(\frac{D}{J}\right)^2 \frac{\zeta}{\boldsymbol\alpha}\cdot\frac{|\nabla T| \cdot l_m}{T}\sim \left(\frac{D}{J}\right)^2\cdot\frac{|\nabla T| \cdot l_m}{T},\label{eq:TNDD_estimate}
\end{align}
in the limit of a weak spin-orbit coupling. Here $\boldsymbol\alpha\approx 1/137$ is the fine-structure constant and we used $\zeta\sim 10^{-2}\sim\boldsymbol\alpha$ in typical transition metal Mott insulators\cite{PhysRevB.78.024402}. Notice that in the absence of spin-orbit coupling, TNDD vanishes since the spin magnetic moment $M$ is a spin-triplet\footnote{We only consider the contribution from the spin magnetic moment in this paper. The orbital magnetic moment in a Mott insulator is a spin-singlet but is much smaller than the spin magnetic moment, by a factor of $(t/U)^2$ in the $(t/U)$-expansion.\cite{PhysRevB.73.155115,PhysRevB.78.024402}}. $D$ and $J$ are the Dzyaloshinskii-Moriya(DM) interaction and the exchange interaction respectively. In a system with a strong spin-orbit coupling one may set $D/J\sim 1$, and $TNDD(\omega)$ is proportional to the ratio of the temperature change across $l_m$ and the temperature. To optimize signal, one may choose a large temperature gradient such that $\nabla T \cdot \mathscr{w}\sim T$ where $\mathscr w$ is the linear system size along the $\nabla T$ direction, and $TNDD(\omega)\sim l_m/\mathscr w$. For instance, $l_m$ of magnetic excitations in a quantum spin ice material was reported to be of the order of a micron\cite{doi:10.7566/JPSJ.87.064702}. For a typical millimeter sample size, $TNDD(\omega)$ can be as large as $10^{-3}$, well detectable within the currently available experimental technology.

\textbf{Crystal symmetry analysis}
TNDD can be phenomenologically described by a tensor $\eta$:
\begin{align}
 \delta_{\hat n}\alpha(\omega)=\sum_{a,b,c}\eta_{abc}(\omega)\hat{\mathcal{E}}_a \hat{\mathcal{B}}_b \nabla_c T\label{eq:phenomenological}
\end{align}
The symmetry condition for $\eta_{abc}(\omega)$ is determined by the fusion rule of two vectors ($\hat{\mathcal{E}},\nabla T$) and one pseudovector ($\hat{\mathcal{B}}$) into a trivial representation under the point group. For any point group, symmetry always allows nonzero $\eta_{abc}$: one may always consider the case $\hat n\sim \hat{\mathcal{E}}\times \hat{\mathcal{B}}$ to be parallel to $\nabla T$. 

As an example, we find that there are four independent response coefficients for the $D_{3d}$ point group:
\begin{align}
 &\delta_{\hat n}\alpha=\eta_1 \nabla_z T (\hat{\mathcal{E}}\times\hat{\mathcal{B}})_z+\eta_2 \hat{\mathcal{B}}_z (\hat{\mathcal{E}}\times \nabla T)_z+\eta_3 \hat{\mathcal{E}}_z (\hat{\mathcal{B}}\times \nabla T)_z\notag\\
 &\;\;+\eta_4 \big[(\hat{\mathcal{E}}_x \hat{\mathcal{B}}_y+\hat{\mathcal{E}}_y \hat{\mathcal{B}}_x)\nabla_y T-(\hat{\mathcal{E}}_x \hat{\mathcal{B}}_x-\hat{\mathcal{E}}_y \hat{\mathcal{B}}_y)\nabla_x T\big]\label{eq:D3d}
\end{align}
Here the x-axis is a $C_2$-axis and the yz-plane is a $\sigma_d$ mirror-plane in the $D_{3d}$ group. The $D_{3d}$ point group is realized in the QSL candidate Herbertsmithite, in the Heisenberg model on the Kagome lattice with DM interactions (see below and Fig.\ref{fig:DM_terms}(b)), as well as in the generalized Kitaev-Heisenberg model on the honeycomb lattice\cite{PhysRevLett.102.017205,PhysRevLett.105.027204,Takagi2019}, relevant for Na$_2$IrO$_3$\cite{PhysRevB.82.064412} and RuCl$_3$\cite{PhysRevB.91.144420,PhysRevB.92.235119,PhysRevB.93.134423}.

\textbf{Microscopic model} We present a concrete microscopic calculation for the TNDD spectrum. The nearest neighbor spin-1/2 Hamiltonian under consideration is on the kagome lattice:
\begin{align}
 H=J\sum_{<ij>}\vec S_i \cdot \vec S_j+\sum_{<ij>}\vec D_{ij} \cdot \vec S_i\times \vec S_j\label{eq:ham},
\end{align}
This model is relevant for various QSL candidate materials such as ZnCu$_3$(OH)$_6$Cl$_2$ (Herbertsmithite) and Cu$_3$Zn(OH)$_6$FBr, and respects both $\mathcal{T}$ and $\mathcal{I}$. Based on the $D_{3d}$ crystal symmetry for the kagome plane, the DM vector $\vec D_{ij}=-\vec D_{ji}$ has two independent coupling constants: $D_z$ (out-of-plane) and $D_p$ (in-plane)\cite{PhysRevB.66.014422} (see Fig.\ref{fig:DM_terms}(b)). Precisely speaking:
\begin{align}
 \vec D_{ij}=d_{ij}\cdot (D_z\cdot \hat z  +  D_p \cdot \hat z \times \hat{r}_{ij}),
\end{align}
where $d_{ij}=\pm 1$, $\hat{r}_{ij}$ is the unit vector along the direction from the site-$j$ to the site-$i$. As shown in Fig.\ref{fig:DM_terms}(b), in each bow-tie: $
 d_{12}=d_{23}=d_{31}=1,d_{34}=d_{45}=d_{53}=-1.
$

Dipole-coupling with an external electric field $\delta H=-\vec E\cdot \vec P$, the electric polarization $\vec P$ has the following form for the nearest neighbor terms\cite{PhysRevB.87.245106}\footnote{Generally the polarization operator contains spin-triplet terms similar to DM interactions. Here for simplicity we only consider spin-singlet terms which dominate in the weak spin-orbit coupling limit.}:
\begin{align}
P_y=&\frac{\zeta e a}{\sqrt{3}}[\vec S_3\cdot(\vec S_2+\vec S_1-\vec S_5-\vec S_4)-2\vec S_1\cdot\vec S_2+2\vec S_5\cdot \vec S_4],\notag\\
P_x=& \zeta  e  a \cdot [\vec S_3\cdot(\vec S_2-\vec S_1+\vec S_5-\vec S_4)],
\label{eq:polarization}
\end{align}
where $e<0$ is the electron charge, $a$ is the nearest neighbor distance, and $\zeta$ is a dimensionless coupling constant (in this paper $\vec S=\vec \sigma/2$.) $\zeta$ can be generated via a $t/U$ expansion in a Hubbard model\cite{PhysRevB.37.9753}. In the leading order $J=\frac{4t^2}{U}$ and $\zeta=\frac{12t^3}{U^3}$\cite{PhysRevB.87.245106}. \footnote{$\zeta$ also receives contribution from the magneto-elastic coupling. For a typical transition metal Mott insulator, this contribution to polarization is similar in size as the contribution from the $t/U$-expansion\cite{PhysRevB.78.024402,PhysRevB.87.245106}.}

\textbf{$Q_1=Q_2$ $Z_2$ spin liquid: Schwinger boson mean-field treatment} There are extensive numerical evidences that the Heisenberg model on the kagome lattice may realize a QSL ground state, although the nature of the QSL is under debate\cite{PhysRevLett.98.117205,Yan1173,PhysRevLett.109.067201,PhysRevB.87.060405,PhysRevLett.118.137202,PhysRevX.7.031020}. The present work does not attempt to resolve this long-standing puzzle. Instead, we will focus on one candidate spin liquid state, which may be realized in the model Eq.(\ref{eq:ham}): Sachdev's $Q_1=Q_2$ $Z_2$ QSL\cite{PhysRevB.45.12377}. The $Q_1=Q_2$ QSL is a gapped state and can be described using the Schwinger boson mean-field theory\cite{PhysRevB.38.316,PhysRevLett.66.1773,doi:10.1142/S0217979291000158}, in which spin is represented by bosonic spinons:
$
 \vec S_i=\frac{1}{2}b^{\dagger}_{i\alpha}\vec\sigma_{\alpha\beta}b_{i\beta},
$
while boson number per site is subject to the constraint $
 b_{i\alpha}^{\dagger}b_{i\alpha}=2S.$ We then do the usual mean-field decoupling and diagonalize the quadratic mean-field spinon Hamiltonian to obtain three spinon bands. We treat DM interaction as a perturbation and keep contributions up to the linear order of $D/J$. Under this approximation we arrive at the following mean-field Hamiltonian.
\begin{align}
 H_{MF}&=-\mu\sum_i(b_{i\alpha}^{\dagger}b_{i\alpha}-2S )-\frac{J}{2}\sum_{\langle ij\rangle}( A_{ij}^*\hat A_{ij}+h.c.)\notag\\
&+\sum\limits_{\langle ij\rangle}(\frac{\vec D_{ij}}{4}\cdot  A_{ij}\hat{\vec C}^{\dagger}_{ij}+h.c.),\label{eq:MF_ham}
\end{align}
where operators $\hat{A}_{ij}\equiv b_{i\alpha}\epsilon_{\alpha\beta}b_{j\beta}$ and $\hat{C}_{ij}\equiv -i b_{i\alpha}(\epsilon\vec{\sigma})_{\alpha\beta}b_{j\beta}$. $H_{MF}$ may be viewed as an ansatz to construct variational spin-liquid wavefunctions with parameters $A_{ij},\mu$. In Sachdev's $Q_1=Q_2$ state, $A_{ij}$ have the following spatial pattern: $A_{ij}=d_{ij} A$, and $A$ can be chosen to be real. See Appendix.~\ref{app:details} for more details.

\begin{figure}
 \includegraphics[width=0.4\textwidth]{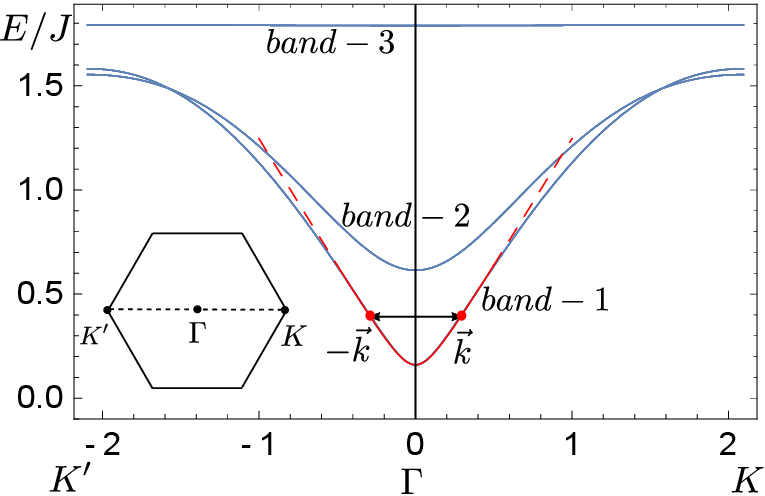}
 \caption{The Schwinger boson band dispersion (blue solid lines) for the mean-field Hamiltonian Eq.(\ref{eq:MF_ham}) of Sachdev's $Q_1=Q_2$ $Z_2$ QSL with parameters $A=1$, $D_z=D_p=0.1J$, and $\mu=-1.792J$. The low energy band-1 near the $\Gamma$ point is well described by the relativistic dispersion Eq.(\ref{eq:relativistic_dispersion}) with gap $\Delta=0.16J$ (red line). The two-spinon (red dots at $\pm \vec k$) contribution to the TNDD response computed in Eq.(\ref{eq:all_temp}) and App.\ref{app:details} is illustrated.}
 \label{fig:bands}
\end{figure}

After Bogoliubov diagonalization, three bands are found: $H_{MF}=\sum_{\vec k,u=1,2,3}^{\alpha=\uparrow,\downarrow} E_{u,\vec k}\gamma^{\alpha\dagger}_{u,\vec k}\gamma_{u,\vec k}^{\alpha}$ as shown in Fig.\ref{fig:bands}, where $\uparrow,\downarrow$ label the Kramers degeneracy. Tuning chemical potential $\mu$ so that the band structure is near the boson condensation at $\Gamma$, the lowest energy band $u=1$ is well described by a relativistic boson disperson:
\begin{align}
 E_{1,\vec k}\approx \sqrt{\Delta^2+\hbar^2k^2v^2},\label{eq:relativistic_dispersion}
\end{align}
where $\Delta$ is the bosonic spinon gap.

\textbf{TNDD contributed from the bosonic spinons}
In the low temperature limit, the two-spinon contribution dominates TNDD with  $\ket{f}\sim \gamma_{u,\vec{k}}^{\alpha,\dagger}\gamma_{v,-\vec{k}}^{\beta\dagger} \ket{i}$ in Eq.(\ref{eq:tndd_micro}). \footnote{Notice that a single spinon excitation is not gauge invariant and does not contribute to physical responses}. $\rho_i,\rho_f$ in Eq.(\ref{eq:tndd_micro}) is related to the nonequilibrium bosonic spinon occupation $g_{u,\vec k}$. From a simple Boltzmann equation analysis with a single relaxation time $\tau$, $g_{u,\vec k}$ deviates from the equilibrium occupation $g^0_{u,\vec k}=\frac{1}{e^{\beta (\vec r)E_{u,\vec k}}-1}$ by $\delta g_{u,\vec k}= \frac{\partial g^0_{u,\vec k}(\vec r)}{\partial E} E_{u,\vec k} \frac{\tau \vec v_{u,\vec k} \cdot \nabla T}{T(\vec r)}$, where $\vec v_{u,\vec k}=\frac{\partial E_{u,\vec k}}{\hbar \partial \vec  k}$. This $\delta g_{u,\vec k}$ is responsible for TNDD.

\begin{figure}
 \includegraphics[width=0.4\textwidth]{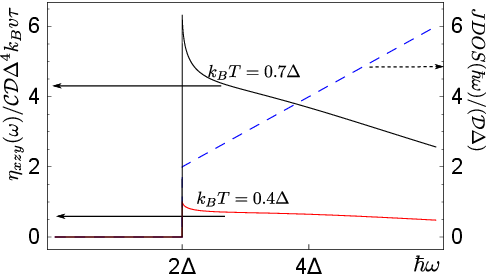}
 \caption{The bosonic two-spinon contribution to TNDD spectra of Sachdev's $Q_1=Q_2$ $Z_2$ QSL Eq.(\ref{eq:MF_ham}) at the temperature $k_B T=0.7\Delta$ (solid black line) and $k_B T=0.4\Delta$ (solid red line), together with the two-spinon joint density of states (dashed blue line).}
 \label{fig:TNDD_curve}
\end{figure}

Since TNDD is a bulk response we consider a 3D system consisting of stacked 2D layers each described by the model Eq.(\ref{eq:ham}) with an interlayer distance $d$. Using the electric polarization Eq.(\ref{eq:polarization}) and spin magnetic moment $\vec M=g_s\mu_B \vec S$, in App.(\ref{app:details} we compute the low temperature/energy TNDD response tensor defined in Eq.(\ref{eq:phenomenological}) within our mean-field treatment (corresponding to $\eta_2$ in Eq.(\ref{eq:D3d})). As plotted in Fig.\ref{fig:TNDD_curve}, we find that ($x,y,z$-directions are illustrated in Fig.\ref{fig:DM_terms})
\begin{align}
 &\eta_{xzy}(\omega)=\mathcal{C}\cdot  \big[1+2 g^0(\hbar\omega/2)\big]\cdot (k_B T)^3\notag\\
 &\cdot\left[3 G_3( z)-3\mbox{ln}z \cdot G_2( z)+(\mbox{ln}z)^2 G_1(z) \right]\notag\\
 &\cdot e^{-\sqrt{(\hbar\omega/2)^2-\Delta^2}/\Delta}\cdot\hbar\omega\cdot JDOS(\hbar\omega)\cdot \frac{\tau\cdot v}{T}.\label{eq:all_temp}
\end{align}
Here the constant $\mathcal{C}\equiv 8\pi\mathscr u_0\boldsymbol\alpha^2 \zeta a a_0\cdot\frac{ \mu_r g_s a^3}{\hbar^3 v^3 }$, where $a_0$ is the Bohr radius. $\mathscr u_0\propto (D/J)^2$ is a dimensionless constant related to the mean-field band structure and can be determined numerically. For the parameters $D_z=D_p=0.1J$ and $\mu=-1.792J$ we find that $\mathscr u_0=0.603$. The 3D optical joint density of states $JDOS(\hbar\omega)\equiv \mathcal{D}\cdot\hbar\omega\cdot\Theta(\hbar\omega-2\Delta)$ where $\mathcal{D}\equiv \frac{1}{8\pi\hbar^2 v^2 d}$.
$g^{0}(\hbar\omega/2)=\frac{1}{e^{\hbar\omega/2k_B T}-1}$, $z\equiv e^{-\frac{\Delta}{k_B T}}$, and $G_\nu (z)\equiv\frac{1}{\Gamma(\nu)}\int_0^{\infty}\frac{x^{\nu-1}dx}{z^{-1}e^x-1}$ is the Bose-Einstein integral. Eq.(\ref{eq:all_temp}) holds when the temperature and the photon energy are within the regime of the relativistic dispersion Eq.(\ref{eq:relativistic_dispersion}). 

In the limit $k_B T\ll \Delta$, Eq.(\ref{eq:all_temp}) can be simplified and we have $\eta_{xzy}(\omega)\propto e^{-\Delta/k_BT}$, where the thermal activation factor can be traced back to $\delta g_{\vec k}$. Importantly, beyond the mean-field treatment, TNDD is generally $\propto \delta \rho_i,\delta\rho_f\propto \nabla T\cdot \tau $ in Eq.(\ref{eq:tndd_micro}), and a thermal activation factor $e^{-\Delta/k_B T}$ in TNDD is always due to the energy diffusion near the mobility gap $\Delta$. Therefore TNDD can serve as a sharp measurement of the \emph{mobility gap} $\Delta$ of the magnetic excitations.

\textbf{Discussion} \textit{Bosonic vs. fermionic spinons}
We computed the TNDD response contributed from bosonic spinons in the Sachdev's $Q_1=Q_2$ $Z_2$ QSL. Fermionic spinons also exist in this $Z_2$ QSL and their contribution to TNDD can be similarly computed in a dual Abrikosov fermion approach\cite{PhysRevB.83.224413,PhysRevB.96.205150}. Without pursuing this calculation in details, one expects that the bosonic factor $[1+2g^0(\hbar\omega/2)]$ (Bose-Einstein integrals) in Eq.(\ref{eq:all_temp}) will be replaced by the corresponding fermionic factor $[1-2f^0(\hbar\omega/2)]$ (Fermi-Dirac integrals), where  $f^0(\hbar\omega/2)=1/(e^{\hbar\omega/2k_B T}+1)$. The contributions from the bosonic spinons and fermionic spinons have different temperature dependence, which, in principle, may be used to detect the statistics of quasiparticles in certain situations.

\textit{Magnetically ordered states}
It is also interesting to consider the TNDD response in a conventional magnetically ordered state respecting either $\mathcal{I}$, or $\mathcal{T}$ combined with a lattice-translation symmetry (as in the case of an antiferromagnet), or both. One may similarly consider the two-magnon contribution to the TNDD response, which probes the joint density of states of magnons. Our estimate Eq.(\ref{eq:TNDD_estimate}) will be modified as follows (see Appendix \ref{app:dm} for details). If the magnetic order is non-collinear, which breaks spin-rotational symmetry completely, the $(D/J)^2$ factor in  Eq.(\ref{eq:TNDD_estimate}) is replaced by $\sim 1$. If the magnetic order is collinear, which only breaks the spin-rotation symmetry down to $U(1)$, the $(D/J)^2$ factor is replaced by $D/J$.

\textbf{Conclusion} In this paper we propose the temperature-gradient-induced nonreciprocal directional dichroism (TNDD) spectroscopy experiment in Mott insulators. Comparing with traditional probes for magnetic excitations, TNND spectroscopy has unique advantages: it directly probes mobile magnetic excitations and decouples from local impurity modes and phonon modes. For instance, an activation behavior $\propto e^{-\Delta/k_B T}$ in the temperature dependence of TNDD sharply measures the \emph{mobility gap} $\Delta$ of the magnetic excitations, a quantity challenging to measure using traditional probes but of fundamental importance in the field of topologically ordered QSL.

The present work can be viewed as one example in a large category of nonlinear thermo-electromagnetic responses. There are other interesting effects. For instance, a temperature gradient also induces a circular dichroism in a system respecting both $\mathcal T$ and $\mathcal I$. We leave these other responses as topics of future studies. 

We thank Kenneth Burch and Di Xiao for helpful discussions. XY and YR acknowledge support from the National Science Foundation under Grant No. DMR-1712128.

\appendix
\section{Localized modes}\label{app:localized_modes}
Let us consider the situation of a Mott insulator in the presence of impurities/disorders, which could introduce localized magnetic modes. Below we consider the contribution to TNDD response from these localized modes. 

Firstly, we comment on the meaning of ``localized modes" discussed here. In an isolated localized phase of matter, like a many-body localized phase(see Ref.\cite{doi:10.1146/annurev-conmatphys-031214-014726,RevModPhys.91.021001} for reviews), thermalization breaks down and the meaning of a temperature-gradient is unclear. We are \emph{NOT} discussing the TNDD response in this situation.

In realistic quantum materials, the magnetic localized modes are coupled with a thermal bath (e.g., phonon thermal bath) and a local temperature is well defined. To facilitate the discussion, one may consider a system with a $U(1)$ spin rotation symmetry in order to sharply define a magnetic localized mode. In addition, we assume a finite mobility gap $\Delta$ of the $U(1)$ charge, and magnetic localized excitations may exist below $\Delta$. Assuming $l_m$ being the mean-free path for mobile magnetic excitations, practically the localized magnetic modes may fall into two regimes according to the localization length $\xi$:

\textbf{(1): $\xi\ll l_m$.} This is the more common situation realized in practical materials. Here the localized magnetic modes may be extrinsic magnetic impurity atoms, or may form at crystalline defects. They may also form at the centers of the vortices of valence bond solid (VBS) order\cite{PhysRevB.70.220403}. Typically the localization length $\xi$ of these magnetic modes is of the same order as the lattice spacing $a$, while $l_m\gg a$ in a reasonably clean Mott insulator. 

It is difficult to model a magnetic localized mode with $\xi\sim a$ since lattice scale details cannot be neglected. Instead, we consider the following situation $a\ll\xi\ll l_m$ so that a low energy effective description is still valid. As a crude model for such magnetic localized modes, one may consider a quantum dot of size $\xi$ in the presence of a temperature gradient; for instance, the left (right) edge of the quantum dot is in contact with a heat reservior at temperature $T_L$ ($T_R$). The modes in the quantum dot are travelling ballistically since $\xi\ll l_m$. Consequently the right-mover (left-mover) in the quantum dot is at temperature $T_L$ ($T_R$). Such a nonequilibrium ensemble is quantitatively comparable with a large (energy-)diffusive system in the presence of the same temperature gradient but with $l_m\sim \xi$ (for example, see Eq.(\ref{eq:delta_g})). Namely, in the present situation, $\xi$ replaces the role of $l_m$ in our estimate Eq.(\ref{eq:TNDD_estimate}). we conclude that the dimensionless ration $TNDD(\omega)$ contributed by such localized modes is reduced by a factor of $\sim \xi/l_m$.

\textbf{(2): $\xi\gg l_m$.} In this situation, the system hosts would-be mobile modes. These modes scatter with disorder multiple times before eventually become localized. For instance, Anderson weak-localization in two spatial dimensions happens with $\xi$ parametrically larger than $l_m$. It is instructive to consider a system size $L$ satisfying $\xi>L>l_m$. For such a system size the localization physics is not present yet. Because photon absorption is still a local process, we expect that the contribution to the TNDD response from such localized modes to be comparable with that from mobile modes.

In summary, the contribution to TNDD response from localized modes in the regime $\xi\ll l_m$ can be safely neglected. In the opposite regime $\xi\gg l_m$, the localized modes still contribute to TNDD significantly. Nevertheless, the localized modes in the latter regime are would-be extended (mobile) states in the absence of disorder.

\section{Spin-orbit coupling and the estimate of TNDD response}\label{app:dm}
From the discussion in the main text and Eq.(\ref{eq:tndd_micro}), up to matrix element effects, the TNDD spectroscopy directly probes the joint density of states $JDOS(\hbar\omega)$ of the mobile magnetic excitations:
\begin{align}
 \delta_{\hat n}\alpha(\omega)\equiv \alpha_{\hat n}(\omega)-\alpha_{-\hat n}(\omega)\propto \hbar\omega\cdot JDOS(\hbar\omega)\cdot \nabla T\cdot \tau\label{eq:TNDD}
\end{align}

In order to estimate the optical absorption coeffient $\alpha_{\hat n}$ in a Mott insulator, one need to estimate the strength of electric polarization and the magnetic dipole moment. It turns out that they are comparable in a typical transition metal Mott insulator, which is drastically different from the case of a band metal/insulator. In the latter case the electric polarization carried by a typical particle-hole excitation is $\sim e\cdot a$ where $e$ is the electron charge and $a$ is the lattice constant, while the magnetic moment carried by the same excitation is of the order of a Bohr magneton $\mu_B$. For a given electromagnetic wave, the magnetic dipole energy scale $\mu_B \cdot B$ is smaller than the electric dipole energy scale $e\cdot a\cdot E$  by roughly a factor of the fine-structure constant $\sim 1/137$, which is why the magnetic dipole processes are often neglected in a band metal/insulator.

In a Mott insulator, however, the electric polarization carried by a magnetic excitation is heavily reduced. In the framework of the Hubbard model, this electric polarization can be estimated as $\zeta \cdot e\cdot a$ where the dimensionless factor $\zeta\sim 8(t/U)^3$\cite{PhysRevB.78.024402}. On the other hand, the magnetic dipole moment carried by the same excitation is still $\sim \mu_B$. As a result, they would have comparable sizes for typical 3d transition metal Mott insulators with $t/U\sim 10$.

The absorption coefficient due to the electric dipole processes can be estimated based on the Fermi's golden rule:
\begin{align}
 \alpha_{\hat n}(\omega)&\sim\frac{2}{n_r\epsilon_0 c}\frac{2\pi}{\hbar} \cdot |\langle f|P|i\rangle|^2 \cdot \hbar\omega\cdot JDOS(\hbar\omega)\notag\\
 &\sim \frac{16\pi^2}{n_r}\boldsymbol\alpha \zeta^2 a^2 \cdot \hbar\omega\cdot JDOS(\hbar\omega).\label{eq:B2}
\end{align}
where $n_r$ is the relative refractive index of the material, $c$ is the speed of light, $\boldsymbol{\alpha}$ is the fine structure constant $\sim 1/137$, and $JDOS(\hbar\omega)$ is the joint density of states for the relevant excitations at photon energy $\hbar\omega$. We assume that the temperature is comparable with the magnetic excitation energy scale, and we have used the typical matrix element $\langle f|P|i\rangle\sim \zeta \cdot e\cdot a$ where $a$ is the lattice constant.

Notice that $JDOS(\hbar\omega)$ may be estimated as $\sim \frac{1}{a^3}\frac{1}{W}$ where $a$ is the lattice constant and $W$ is the band width of the excitations. For a typical photon energy $\sim W$, one finds that $\hbar\omega\cdot JDOS(\hbar\omega)\sim 1/a^3$, independent of the nature of the excitations. For instance, the interband absorption coefficient $\alpha(\omega)$ in a band metal/insulator is typically $\sim 10^7 \mbox{m}^{-1}$. The dimensionless coupling constant $\zeta$ reduces by a factor of $~10^2$ in transition metal Mott insultors, which gives the absorption coefficient $\sim 10^3\mbox{m}^{-1}$, broadly consistent with the tera-Hertz penetration depth ($\sim 1$mm) for these quantum magnets\cite{PhysRevLett.111.127401,PhysRevLett.119.227201}.

The TNDD response can be similarly estimated. We first consider the case of a quantum paramagnet.
\begin{align}
 \delta_{\hat n}\alpha(\omega)&\sim\frac{2\mu_r}{\epsilon_0 c^2}\frac{2\pi}{\hbar} \cdot (\rho_i-\rho_f)\cdot 2 \cdot \mbox{Re}[\langle f|P|i\rangle\langle i|M|f\rangle]\notag\\
 &\;\;\;\;\;\cdot \hbar\omega\cdot JDOS(\hbar\omega)
\end{align}
We again assume that the temperature is comparable with the magnetic excitation energy scale, and consequently the effect of temperature gradient in $(\rho_i-\rho_f)$ can be estimated by the dimensionless factor $\frac{|\nabla T|\cdot l_m}{T}$ where $l_m$ is the mean-free path of the magnetic excitations. 
\emph{If the spin-orbit coupling (SOC) is strong} one may estimate $\langle f|P|i\rangle\sim \zeta e a$ while $\langle i|M|f\rangle\sim g_s\mu_B$ ($g_s$ is the g-factor the spin magnetic moment.). Putting together we have:
\begin{align}
 \delta_{\hat n}\alpha(\omega)\sim 16\pi^2 \mu_r g_s \boldsymbol\alpha^2 \zeta a_0 a&\cdot\frac{|\nabla T|\cdot l_m}{T}\cdot\hbar\omega\cdot JDOS(\hbar\omega),\notag\\
 &\mbox{ if strong SOC.}\label{eq:B4}
\end{align}
Here $a_0$ is the Bohr radius.

From Eq.(\ref{eq:B2},\ref{eq:B4}), and $a_0\sim a$, we can estimate that if the spin-orbit coupling is strong and the temperature is comparable with the magnetic excitation energy scale, the dimensionles ratio $TNDD$ in Eq.(\ref{eq:TNDD_ratio})
\begin{align}
 TNDD(\omega)\sim \frac{\boldsymbol\alpha}{\zeta}\frac{|\nabla T|\cdot l_m}{T}\sim \frac{|\nabla T|\cdot l_m}{T} ,\mbox{ if strong SOC.}
\end{align}
Here we used the fact that for a typical transition metal Mott insulator $\zeta\sim 10^{-2}\sim \boldsymbol\alpha$.

In the absence of the SOC, $\langle i|M|f\rangle=0$ because $\vec M =g_s\mu_B \vec S$  is proportional to the conserved total spin $\vec S$ (We only consider the spin magnetic moment. The orbital magnetic moment in Mott insulators is much smaller and neglected.). In the limit of a weak SOC: $D/J\ll 1$, the TNDD response can be estimated as follows. The only effect of the weak SOC is in the matrix element product: $\langle f|P|i\rangle\langle i|M|f\rangle$. 

For the magnetic dipole matrix element: $\langle i|M|f\rangle \propto \frac{1}{E_f-E_i}\langle i|[S,H]|f\rangle\propto \langle f|\vec D \cdot [S ,\vec S_i\times \vec S_j]|i\rangle$. Notice that the operator of the commutator is a spin triplet. There are two possibilities: (1): the states $|f\rangle$ and $|i\rangle$ differ by spin-1 in the limit $D/J\rightarrow 0$. For instance, $|f\rangle$ may be a spin triplet while $|i\rangle$ is a spin singlet in that limit; (2): the states $|f\rangle$ and $|i\rangle$ have the same spin in the limit $D/J\rightarrow 0$.

In the situation-(2), the magnetic dipole matrix element $\langle i|M|f\rangle \propto (D/J)^2$, because the wavefunction corrections of $|f\rangle$ and $|i\rangle$ due to nonzero $D/J$ need to be considered. In this situation, the electric dipole matrix element $\langle f|P|i\rangle\propto (D/J)^0$ since $P$ is a spin singlet operator in the limit of $D/J\rightarrow 0$. Therefore in situation-(2) we have $\langle f|P|i\rangle\langle i|M|f\rangle\propto (D/J)^2$.

In the situation-(1), a similar consideration leads to: $\langle i|M|f\rangle \propto (D/J)$ and $\langle f|P|i\rangle\propto (D/J)$. So we still have $\langle f|P|i\rangle\langle i|M|f\rangle\propto (D/J)^2$. 

In summary, we have the following estimate in a quantum paramagnet assuming the temperature is comparable with the magnetic excitation energy scale:
\begin{align}
 TNDD(\omega)&\sim \left(\frac{D}{J}\right)^2\frac{\boldsymbol\alpha}{\zeta}\frac{|\nabla T|\cdot l_m}{T}\notag\\
 &\sim \left(\frac{D}{J}\right)^2\frac{|\nabla T|\cdot l_m}{T} ,\mbox{ if weak SOC.}\label{eq:app_estimate}
\end{align}

Next we estimate the TNDD response in magnetic ordered states due to magnon excitations. Even in the absence of microscopic SOC, the $(D/J)^2$ factor in the estimate Eq.(\ref{eq:app_estimate}) will be replaced by $\sim 1$ in a non-collinear magnetic ordered state, because the spin-rotation symmetry is completely broken. 

In a collinear magnetic ordered state, the spin rotation symmetry is broken down to $U(1)$ in the absence of SOC. The electric polarization operator $P$ is expected to carry zero charge under this $U(1)$ rotation. To have a nonzero matrix element product $\langle f|P|i\rangle\langle i|M|f\rangle$, one must consider the linear-order effect of the SOC. Therefore in this case the $(D/J)^2$ factor in the estimate Eq.(\ref{eq:app_estimate}) will be replaced by $\sim D/J$.

\section{Details of the mean-field calculation for TNDD}\label{app:details}
In this section we provide a detailed account of the Schwinger boson mean-field theory. The spin is represented by bosonic spinons
\begin{align}
 \vec S_i=\frac{1}{2}b^{\dagger}_{i\alpha}\vec\sigma_{\alpha\beta}b_{i\beta},
\end{align}
while boson number per site is subject to the constraint:
\begin{align}
 b_{i\alpha}^{\dagger}b_{i\alpha}=\kappa.\label{eq:kappa}
\end{align}
Although $\kappa=2S$ for spin-$S$, it will be convenient to consider $\kappa$ to be a continuous parameter, taking on any non-negative value\cite{PhysRevB.45.12377,PhysRevB.74.174423}.

Considering the operator identities $\vec S_i\cdot \vec S_j=-\frac{1}{2}\hat A_{ij}^{\dagger}\hat A_{ij}+\frac{\kappa^2}{4}$ and $ \vec S_i\times \vec S_j=\frac{1}{4}[ \hat{\vec C}_{ij}^{\dagger}\hat A_{ij}+h.c.]$, where
\begin{align}
 \hat A_{ij}&=-\hat A_{ji}=b_{i\alpha}\epsilon_{\alpha\beta}b_{j\beta},&\hat{\vec{C}}_{ij}&=\hat{\vec C}_{ji}=-\ri b_{i\alpha}(\epsilon\vec \sigma )_{\alpha\beta}b_{j\beta}.
\end{align}
standard mean-field decoupling of Eq.(\ref{eq:ham}) leads to the mean-field Hamiltonian:
\begin{align}
 H_{MF}&=-\frac{J}{2}\sum_{<ij>} (A_{ij}^*\hat A_{ij}+A_{ij}\hat A_{ij}^{\dagger}-|A_{ij}|^2)\notag\\
 &+\sum_{<ij>}\frac{\vec D_{ij}}{4}\cdot (\vec C_{ij}^* \hat A_{ij}+ A_{ij}\hat{\vec C}^{\dagger}_{ij}-\vec C_{ij}^*A_{ij}+h.c.)\notag\\
 &-\mu\sum_i(b_{i\alpha}^{\dagger}b_{i\alpha}-\kappa ).\label{eq:HMF}
\end{align}
Here the chemical potential $\mu$ is introduced to enforce constraint Eq.(\ref{eq:kappa}) on the mean-field level. $H_{MF}$ may be viewed as an ansatz to construct variational spin-liquid wavefunctions with parameters $A_{ij},\vec C_{ij},\mu$. 

We will consider the case of a small $D/J$ and keep contributions up to the linear order of $D/J$. Under this approximation we will set the parameter (\emph{not the operator $\hat{\vec{C}}_{ij}$}) $\vec C_{ij}\propto D/J$ to zero in Eq.(\ref{eq:HMF}) below, which yields Eq.\eqref{eq:MF_ham} in the main text. We also focus on Sachdev's $Q_1=Q_2$ state, where $A_{ij}$ happens to have the following spatial pattern: $A_{ij}=d_{ij} A$, and $A$ is chosen to be real.

After diagonalizing $H_{MF}$ in the momentum space, there are three Kramers degenerate Bogoliubov boson bands (see Fig.\ref{fig:bands}):
\begin{align}
 H_{MF}=\sum_{\vec k,\alpha=\uparrow,\downarrow}^{u=1,2,3} E_{u,\vec k}\gamma_{u,\vec k}^{\alpha \dagger}\gamma_{u,\vec k}^{\alpha}.\label{eq:MF_ham_bogoliubov}
\end{align}
Notice that spin is not a good quantum number and $\uparrow,\downarrow$ are simply labelling the two-fold Kramers degeneracy for each band. 

In the presence of a temperature gradient $\nabla T(\vec r)$, the occupation of Bogoliubov spinons $g_{u,\vec k}=\langle n_{u,\vec k} \rangle $ (where $n_{u,\vec k}=\gamma_{u,\vec k}^{\dagger}\gamma_{u,\vec k}$) deviates from the thermal equilibrium value $g_{u,\vec k}^0$. For simplicity, we consider the steady state Boltzmann equation within a single relaxation-time approximation:
\begin{align}
\vec v_{u,\vec k}\cdot \nabla_{\vec r} g_{u,\vec k}(\vec r) = -\frac{g_{u,\vec k}(\vec r)-g_{u,\vec k}^0(\vec r)}{\tau},
\end{align}
where $g_{u,\vec k}^0(\vec r)=\frac{1}{e^{ E_{u,\vec k}/k_B T(r)}+1}$, $\vec v_{u,\vec k}=\frac{1}{\hbar}\nabla_{\vec k}E_{u,\vec k}$.
%And
%\begin{align}
%\nabla_{\vec r} g^0_{u,\vec k}(\vec r)=\frac{\partial g^0_{u,\vec k}(\vec r)}{\partial E} E_{u,\vec k} \frac{-\nabla T}{T(\vec r)}.
%\end{align}
To the leading order, these give $\delta g_{u,\vec k}(\vec r)\equiv g_{u,\vec k}(\vec r)-g_{u,\vec k}^0(\vec r)$:
\begin{align}
 \delta g_{u,\vec k}(\vec r)\equiv\delta g^{\uparrow}_{u,\vec k}(\vec r)=\delta g^{\downarrow}_{u,\vec k}(\vec r)=\frac{\partial g^0_{u,\vec k}(\vec r)}{\partial E} E_{u,\vec k} \frac{\tau \vec v_{u,\vec k} \cdot \nabla T}{T(\vec r)}\label{eq:delta_g}
\end{align}
Since the velocity $\vec v_{u,\vec k}=-\vec v_{u,-\vec k}$, we have:
\begin{align}
 \delta g_{u,\vec k}(\vec r)=-\delta g_{u,-\vec k}(\vec r)\label{eq:occupation_inversion}
\end{align}

To be concrete, we focus on the case $\hat{\mathcal{E}}=\hat x$ and $\hat{\mathcal{B}}=\hat z$, with the light propagating direction $\hat n=-\hat y$ and the temperature gradient $\nabla T\propto\hat y$(the $\eta_2$ response in Eq.(\ref{eq:D3d})).
In order to compute the matrix elements in Eq.(\ref{eq:tndd_micro}), one writes $P_x$ and $M_z$ in terms of the Bogoliubov bosons, and selects the relevant terms:
\begin{align}
P_x&\rightarrow \sum^{v,\alpha,w,\alpha'}_{\vec q}X^{v,\alpha,w,\alpha'}_{\vec q }\gamma_{v,\vec q}^{\alpha\dagger} \gamma^{\alpha'\dagger}_{w,-\vec q}+h.c.\notag\\
&\;\;\;\;+\frac{1}{A}\sum_{\vec q,\vec p}^{v,\alpha,w,\alpha',t,\beta} Y^{v,\alpha,w,\alpha',t,\beta}_{\vec q,\vec p}\gamma_{v,\vec q}^{\alpha\dagger} \gamma^{\alpha'\dagger}_{w,-\vec q} \gamma_{t,\vec p}^{\beta\dagger} \gamma^{\beta}_{t, \vec p}+h.c.,\notag\\
M_z&=-g_s\mu_b\sum_i b^{\dagger}_{i\alpha}\frac{\sigma^z_{\alpha\beta}}{2}b_{i\beta}\notag\\
&\;\;\;\;\rightarrow\sum^{v,\alpha,w,\alpha'}_{\vec q} Z^{v,\alpha,w,\alpha'}_{\vec q }\gamma_{v,\vec q}^{\alpha\dagger} \gamma^{\alpha'\dagger}_{w,-\vec q}+h.c. .
\end{align}
The objects $X^{v,\alpha,w,\alpha'}_{\vec q}$, $Y^{v,\alpha,w,\alpha',t,\beta}_{\vec q,\vec p}$, $Z^{v,\alpha,w,\alpha'}_{\vec q }$ are determined by the Bogoliubov transformation from Eq.(\ref{eq:MF_ham}) to Eq.(\ref{eq:MF_ham_bogoliubov}).

Plugging in Eq.(\ref{eq:tndd_micro}), one finds 
\begin{align}
\delta_{\hat n}\alpha(\omega)=\frac{8\pi\mu_r}{\epsilon_0 c^2d}\mbox{Re}[I(\omega)],\label{eq:I_integral}
\end{align}
 where
\begin{align}
 &I(\omega)=\frac{\omega}{A}\sum_{\vec q}^{v,\alpha,w,\alpha'}\Big[X^{v,\alpha,w,\alpha'}_{\vec q }+\frac{1}{A}\sum_{\vec p}^{t,\beta}Y^{v,\alpha,w,\alpha',t,\beta}_{\vec q,\vec p}\cdot g_{t,\vec p}\Big]^*\notag\\
 &\cdot Z^{v,\alpha,w,\alpha'}_{\vec q }(1+g_{v,\vec q}+g_{w,-\vec q})\cdot\delta(E_{v,\vec q}+E_{w,-\vec q}-\hbar\omega).\label{eq:dichroism}
\end{align}
Here the bosonic factor $(1+g_{v,\vec q}+g_{w,-\vec q})$ is well anticipated from the golden rule. The factor $g_{t,\vec p}$ appears because of the quartic interactions in $\vec P$ in Eq.(\ref{eq:polarization}).

It is a good moment to study the symmetry property of $I(\omega)$. In thermal equilibrium, it is straightforward to see that the inversion symmetry alone dictates $I(\omega)=0$, while time-reversal symmetry alone allows a nonzero imaginary part of $I(\omega)$ (giving rise to the well-known natural circular dichroism in noncentrosymmetric systems).

Next we consider the effect of nonequilibrium occupation $\delta g_{u,\vec k}$ in Eq.(\ref{eq:delta_g}). Expanding Eq.(\ref{eq:dichroism}) gives three contributions, $I=I^{(A)}+I^{(B)}+I^{(C)}$:
\begin{align}
 I^{(A)}(\omega)&\propto X^*\cdot Z\cdot(\delta g_{v,\vec q}+\delta g_{w,-\vec q}),\notag\\
  I^{(B)}(\omega)&\propto Y^*\cdot Z\cdot g^0_{t,\vec p}\cdot (\delta g_{v,\vec q}+\delta g_{w,-\vec q}),\notag\\
   I^{(C)}(\omega)&\propto Y^*\cdot Z\cdot\delta g_{t,\vec p}\cdot (1+g^0_{v,\vec q}+g^0_{w,-\vec q}).\label{eq:3_contributions}
\end{align}
While the inversion symmetry allows all these contributions, the time-reversal symmetry only allows their real parts: the directional dichroism. In addition, in the special situation that $v=w$, namely if the created two spinons are in the same band, obviously $I^{(A)}(\omega)=I^{(B)}(\omega)=0$ due to Eq.(\ref{eq:occupation_inversion}) and only $I^{(C)}(\omega)$ is nonzero.

\begin{figure}
 \includegraphics[width=0.45\textwidth]{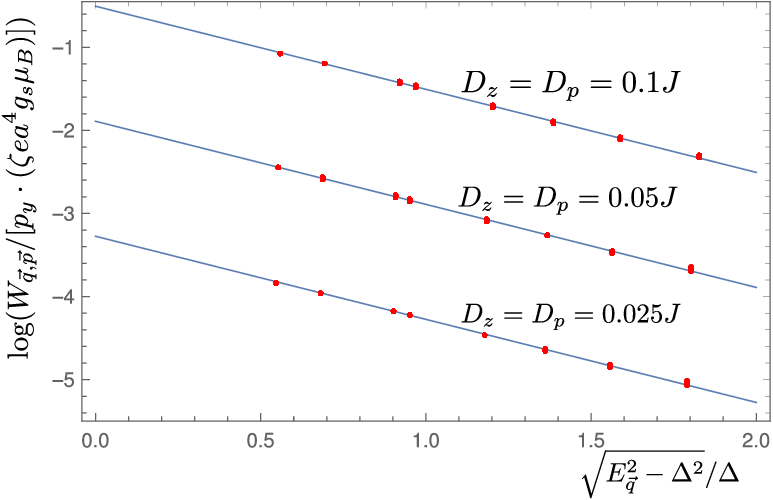}
 \caption{The fit $\mbox{log}(W_{\vec q,\vec p}/[p_y\cdot(\zeta e a^4 g_s\mu_B)])=\mbox{log}(\mathscr u_0)-\sqrt{E_{1,\vec q}^2-\Delta^2}/\Delta$ (i.e., Eq.(\ref{eq:W_all}) with $\vec{\mathscr u}=\mathscr u\hat y=\mathscr u_0 \zeta e a^4 g_s\mu_B \hat y$) with only one fitting parameter $\mathscr u_0$. In each case 696 data points with both $\sqrt{E_{1,\vec p}^2-\Delta^2}/\Delta$ and $\sqrt{E_{1,\vec q}^2-\Delta^2}/\Delta$ between $0.5$ and $1.7$ are plotted. Since many data points are related by the lattice symmetry and/or share the same momentum $\vec q$ (but different $\vec p$), the visibly different data points are much fewer. We set $A=1$, and consider three cases of different SOC strength: case-(a): $D_z=D_p=0.025J$ (and $\mu=-1.752J$); case-(b): $D_z=D_p=0.05J$ (and $\mu=-1.765J$); case-(c) $D_z=D_p=0.1J$ (and $\mu=-1.792J$). Notice that for each case the chemical potential $\mu$ is tuned so that the spinon gap is fixed to be $\Delta=0.16J$. As shown in this figure, we numerically find that $\mathscr u_0=0.0378$ in case-(a), $\mathscr u_0=0.151=0.0378\cdot 3.99$ in case-(b), and $\mathscr u_0=0.603=0.151\cdot 3.99$ in case-(c). The scaling $\mathscr u_0\propto (D/J)^2$ is confirmed.}
 \label{fig:W}
\end{figure}

Focusing on the low temperature/energy TNDD spectroscopy, one may consider the contribution $v=w=t=1$ from the lowest energy band only (see Fig.\ref{fig:bands} for a plot of the band structure), and compute $I(\omega)=I^{(C)}(\omega)$ analytically. In this case:
\begin{align}
& I^{(C)}(\omega)=\frac{\omega}{A^2}\sum_{\vec q,\vec p}W_{\vec q,\vec p}\cdot \delta g_{1,\vec p} (1+2 g^0_{1,\vec q})\delta(2E_{1,\vec q}-\hbar\omega),\notag\\
 &\mbox{where } W_{\vec q,\vec p}\equiv\sum_{\alpha,\alpha',\beta}(Y^{1,\alpha,1,\alpha',1,\beta}_{\vec q,\vec p})^*\cdot Z^{1,\alpha,1,\alpha'}_{\vec q}.\label{eq:IC1}
\end{align}
$W_{\vec q,\vec p}$ is a real function satisfying $W_{\vec q,\vec p}=-W_{-\vec q,-\vec p}$ due to the inversion symmetry. Taylor expanding near the $\Gamma$-point, to the leading order one expects:
$W_{\vec q,\vec p}\approx\vec{\mathscr{u}} \cdot \vec p + \vec{\mathscr{v}} \cdot \vec q$.
In fact, interestingly, we numerically found that $W_{\vec q,\vec p}$ can be well described as 
\begin{align}
 W_{\vec q,\vec p}=(\vec{\mathscr u}\cdot \vec p) \;  e^{-\sqrt{E_{1,\vec q}^2-\Delta^2}/\Delta}\label{eq:W_all}
\end{align}
in the momentum regime where the relativistic dispersion Eq.(\ref{eq:relativistic_dispersion}) holds (see Fig.\ref{fig:W} for details). We do not attempt to analytically justify Eq.(\ref{eq:W_all}) here since it deviates from the main purpose of this paper. Eq.(\ref{eq:IC1},\ref{eq:W_all}) then lead to:
\begin{align}
 &I^{(C)}(\omega)=  \frac{\omega}{A^2} \sum_{\vec p} (\vec{\mathscr{u}} \cdot \vec p \;\delta g_{1,\vec p})\notag\\
 &\;\;\;\cdot \sum_{\vec q} e^{-\sqrt{E_{1,\vec q}^2-\Delta^2}/\Delta} (1+2 g^0_{1,\vec q})\delta(2 E_{1,\vec q}-\hbar\omega).\label{eq:IC2}
\end{align}
Crystal symmetry and dimensional analysis show that $\vec{\mathscr u}=\mathscr u\hat y=\mathscr u_0 \zeta e a^4 g_s\mu_B \hat y$, consistent with the $\eta_2$ response in Eq.(\ref{eq:D3d}). The dimensionless number $\mathscr u_0$ is expect to be $\sim (D/J)^2$ and can be determined numerically (see Fig.\ref{fig:W} for details).

With Eq.(\ref{eq:delta_g},\ref{eq:I_integral},\ref{eq:relativistic_dispersion},\ref{eq:IC2}) the low temperature/energy TNDD response can be computed within our mean-field treatment:
\begin{align}
 &\delta_{\hat y}\alpha(\omega)=\mathcal{C}\cdot  \big[1+2 g^0(\hbar\omega/2)\big]\cdot (k_B T)^3\notag\\
 &\cdot\left[3 G_3( z)-3\mbox{ln}z \cdot G_2( z)+(\mbox{ln}z)^2 G_1(z) \right]\notag\\
 &\cdot e^{-\sqrt{(\hbar\omega/2)^2-\Delta^2}/\Delta}\cdot\hbar\omega\cdot JDOS(\hbar\omega)\cdot \frac{\nabla_y T\cdot\tau\cdot v}{T}.\label{eq:all_temp_1}
\end{align}
This is just the Eq.(\ref{eq:all_temp}) in the main text.

We can apply the estimate in the previous section to the present example as follows. We firstly estimate $\alpha_{\hat n}$ due to the electric dipole processes following the golden rule:
\begin{align}
 &\alpha_{\hat n}(\omega)\sim \frac{2}{n_r \epsilon_0 c} \frac{2\pi}{\hbar} (\zeta e a)^2 [1+2 g^0(\hbar\omega/2)]\hbar\omega\cdot JDOS(\hbar\omega)\notag\\
 &=\frac{16\pi^2}{n_r}\boldsymbol\alpha \zeta^2 a^2 [1+2 g^0(\hbar\omega/2)]\hbar\omega\cdot JDOS(\hbar\omega),\label{eq:alpha_estimate}
\end{align}
where $n_r$ is material's relative refractive index. For the situation with $k_B T\sim J\sim \frac{\hbar v}{a}$ and $\hbar\omega\sim 2\Delta$, Eq.(\ref{eq:all_temp},\ref{eq:alpha_estimate}) give the dimensionless ratio $TNDD(\omega)$ in Eq.(\ref{eq:TNDD_ratio}):
\begin{align}
 TNDD(\omega)\sim \frac{ \boldsymbol\alpha a_0}{\zeta a}\cdot\mathscr u_0 \frac{\nabla_y T\cdot\tau\cdot v}{T},
\end{align}
confirming the estimate Eq.(\ref{eq:TNDD_estimate}) since $\mathscr{u}_0\propto (D/J)^2$.

Finally, we would like to remark on the validity of the mean-field treatment. Although we performed the calculation within the mean-field approach, the main component of the calculation (Eq.(\ref{eq:3_contributions},\ref{eq:IC1}) in App.\ref{app:details}) is justified as long as the quasiparticle description is valid. These microscopic contributions to TNDD can be written down phenomenologically as a low quasiparticle-density expansion, up to the second order $\propto g_{\vec p}\cdot g_{\vec q}$. Some other components of the calculation (e.g., the matrix element behavior Eq.(\ref{eq:W_all}) ) may receive corrections moving beyond the mean-field approximation, but these would not change the result of TNDD response qualitatively.

\bibliography{TNDD}
\end{document}